# Streaming Self-Corrected Dual-Comb Spectrometer

Alexander Eber[1], Christoph Gruber[1], Martin Schultze[1], Birgitta Bernhardt[1], Marcus Ossiander[1,2,*]

[1]Institute of Experimental Physics, Graz University of Technology, Petersgasse 16, 8010 Graz, Austria

[2]Harvard John A. Paulson School of Engineering and Applied Sciences, 9 Oxford Street, Cambridge, Massachusetts 02138, United States

*marcus.ossiander@tugraz.at

## Abstract

We radically simplify coherently averaged dual-comb spectroscopy by introducing a real-time self-correction system: a radio frequency system-on-chip computes each incoming dual-comb interferogram's phase, frequency, and arrival time; calculates changes in the combs' carrier-envelope offset frequency and repetition rate difference; and immediately phase-corrects the incoming interferogram data stream. The algorithm supports up to 0.3 GHz interferogram frequency bandwidth and thus combines fast measurement times (corresponding to high detunings) with broadband optical detection. Using this system, we achieve comb-resolved spectroscopy with Fourier-limited linewidth, coherent averaging over arbitrarily long durations, and high signal-to-noise ratios (e.g., up to 2900 for 150 s averaging time). Iodine and acetylene spectroscopy yield excellent agreement with literature over an optical bandwidth of 10 THz in the visible and near-infrared. Common dual-comb spectroscopy self-correction requires a bright and continuous interferogram train. We lift this requirement by introducing cross-channel correction: the algorithm measures phase fluctuations from a reference channel and predicts and corrects their influence on a signal channel. This enables correcting unstable or intermittent signals (typical, e.g., in field measurements), or low-amplitude signals with amplified phase fluctuations (relevant for nonlinearly upconverted combs). The approach makes instantaneous dual-comb spectroscopy available to everyday applications.

## Introduction

By combining the superior resolution of Fourier transform spectroscopy with rapid acquisition speeds, dual-comb spectroscopy is revolutionizing trace gas monitoring[1–5], molecular fingerprinting & precision spectroscopy[6–9], and hyperspectral imaging[10]. New and maturing frequency comb lasers extend possible applications[11–18], reduce cost, and simplify use, moving dual-comb spectrometers to the verge of widespread commercial adaption.

The enabling concept of dual-comb spectroscopy is to replace the mechanical scanning of conventional Fourier transform spectroscopy with inherent properties of two frequency combs with repetition rates $f_{rep,1} = f_{rep}$ and $f_{rep,2} = f_{rep} + \Delta f_{rep}$[19,20]. When spatially superimposed on a photodiode, their radio frequency beating in time (generated by the repetition rate detuning $\Delta f_{rep}$) yields an interferogram that reveals the lasers' complex spectral amplitude at the position of each comb line after Fourier transformation.

Increasing the signal-to-noise ratio of dual-comb spectrometers adds complexity: relative fluctuations between the two combs' repetition rates or carrier-envelope offset frequencies change subsequent interferograms and prevent coherent averaging. The gold standard solution is measuring and locking both combs' repetition rates and carrier-envelope offset frequencies using nonlinear self-referencing[21,22] or interfering with at least one continuous-wave laser[23,24]. However, this approach requires active feedback to the laser sources and numerous active and passive optical and electronic components. A solution applicable to all lasers is measuring the fluctuations and correcting their effect during data acquisition or post-processing[25–32], however, it still adds many additional components and computing time.

Self-correction algorithms allow coherent averaging by deducing phase, repetition rate, and timing fluctuations from the recorded dual-comb interferograms and correcting them[33] – a numerically expensive but powerful method, provided fluctuations do not exceed reasonable bounds. So far, most implementations do not provide real-time spectroscopic data, limiting the maximum averaging time via the available memory, or require a specialized acquisition card, graphics processing unit, and computer to perform the task[28].

Here, we demonstrate that combining a state-of-the-art, high-repetition-rate, single-cavity dual-comb laser system with data acquisition and processing on a state-of-the-art radio frequency system-on-chip allows real-time phase correction and coherent averaging without measuring the combs' carrier-envelope offset frequencies and repetition rates. The advance yields a drastically simplified, flexible, high-performance dual-comb spectrometer, streaming spectra at multi-kHz rate and 1 GHz optical resolution.

## Experimental Setup and Data Acquisition

Fig. 1a shows our experimental setup. A commercial single-cavity dual-comb laser (K2-1000, K2 Photonics AG) provides two frequency combs sharing the same laser cavity but with two slightly different repetition rates around $f_{rep}$ = 1 GHz. Internal second harmonic generation allows splitting the combs' power between a fundamental beam (wavelength ($\pm e^{-2}$ intensity bandwidth): (1050 ± 11) nm, frequency: (285.5 ± 3.2) THz), and its second harmonic (wavelength: (525 ± 4) nm, frequency: (571.0 ± 4.4) THz). For our demonstrations, we use below-mW average power per comb and $\Delta f_{rep}$ = 20 kHz detuning (repetition rate difference). We overlap both combs using a non-polarizing beamsplitter, pass them through a sample cell, and detect the transmitted power using an amplified photodiode (PD1000-VIS, QUBIG GmbH). After a low-pass filter rejects radiofrequency signals beyond 500 MHz (the repetition rates and distant comb-line beatings), we digitize the stream of interferograms using a 14-bit, 4.9 GSa/s analog-to-digital converter on a radio frequency system-on-chip (RFSoC4x2, Real Digital, LLC using a Zynq Ultrascale+ RFSoC XCZU48DR-2FFVG1517E, Advanced Micro Devices, Inc.).

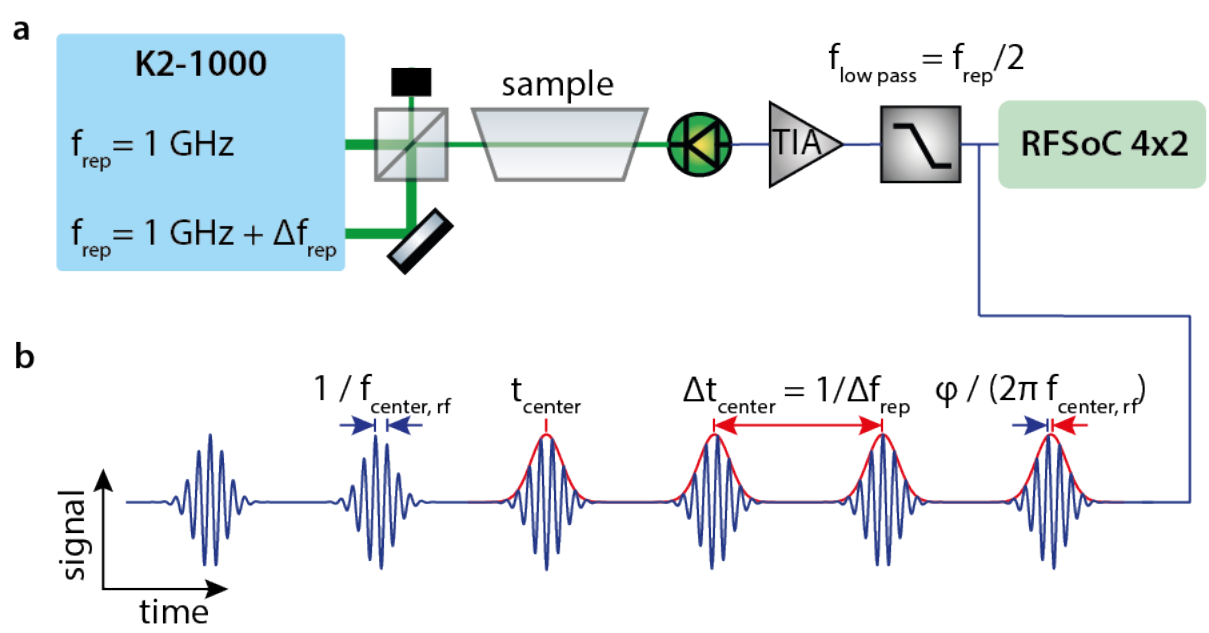

Fig. 1. **Experimental and sampling setup.** a) Two frequency combs generated by a single-cavity dual-comb laser with detuning $\Delta f_{rep}$ are superimposed by a beamsplitter and transmitted through a sample cell. Their beating is detected by an amplified photodiode (TIA) and digitized & processed on a radio frequency system-on-chip (RFSoC). b) Interferograms are detected every $\Delta t_{center}$ and oscillate with frequency $f_{center,rf}$. If the two combs' carrier-envelope offset frequencies fluctuate independently, the interferogram carrier-envelope offset phase $\varphi$ slips by $\Delta\varphi$ between subsequent interferograms.

## Phase and Detuning Fluctuations

Two combs with stable repetition rate detuning $\Delta f_{rep}$ generate an interferogram every $\Delta t_{center} = 1/ \Delta f_{rep}$, visualized in Fig. 1b. If $\Delta f_{rep} > 0$ and both combs' carrier-envelope offset frequencies $f_{ceo\ 1/2}$ are equal, the radio frequency spectrum equals the optical spectrum with all frequencies converted by a factor $m = f_{rep} / \Delta f_{rep}$. I.e., each optical frequency $f_{optical}$ corresponds to a radio frequency $f_{rf} = f_{optical} / m$, and the carrier frequency of the interferogram $f_{center,\ rf} = f_{center,\ optical} / m$ is proportional to the optical carrier frequency $f_{center,\ optical}$. Differing (but stable) carrier-envelope offset frequencies shift the radio frequency spectrum according to $f_{rf} = f_{optical} / m - f_{ceo1} + f_{ceo2} = f_{optical} / m + \Delta f_{ceo}$. However, and in contrast to the optical pulses, the carrier-envelope phase $\varphi$ of the interferogram train (see Fig. 1b) remains constant and the phase slip $\Delta\varphi$ between consecutive interferograms vanishes.

In the time domain, fluctuations of either $f_{ceo\ 1,2}$ introduce a phase slip $\Delta\varphi$ and thus prevent averaging interferograms coherently. In the extreme case, $\Delta\varphi = \pi$, two consecutive interferograms coherently add to zero. Fluctuations further lead to an uncontrollably varying radio frequency-to-optical conversion and thus hamper efficient averaging also in the spectral domain (for data visualizing this effect, see Fig. 3d).

Provided the phase slip $\Delta\varphi$ between two consecutive interferograms is always within [-π, π], $\Delta\varphi$ is deductible from the interferograms themselves, and a self-correction can successfully be applied. The algorithm (described below) measures $\Delta\varphi$ and reveals that the single-cavity dual-comb laser (even though its carrier-envelope offset frequencies are not stabilized) fulfills this criterion (visible in Fig. 3a). Therefore, the phase slip between two consecutive interferograms can be compensated unambiguously. Although we do not stabilize our combs' repetition rates, their detuning changes by less than 2 Hz, consolidating the inherent stability of the laser system and the applicability of a real-time self-correction.

**Phase Correction Algorithm and Laser Stability**

Fig. 2 details the data acquisition and phase correction algorithm: the digitized signal is processed by programmable logic clocked at 307 MHz, i.e., we start with 16 consecutive samples per clock cycle. First, a digital mixer selects the frequency region (up to a 2.3 GHz radio frequency carrier frequency) of interest before we average 8 successive samples, reducing the data rate to 2 real-valued samples per clock cycle. We then calculate the analytical signal using a 51$^{st}$-order finite impulse response Hilbert transform, which diminishes negative frequency components such that we can mix with 307/2 MHz and drop every second sample without losing information. This yields a digital signal arriving as 1 complex-valued sample per clock cycle and supports 307 MHz interferogram radio frequency bandwidth. This is close to the maximally usable interferogram bandwidth ($f_{rep}$ / 2) for most frequency combs and can be extended by raising the fabric clock or processing multiple samples in parallel.

The main design concept of the phase correction algorithm is streaming operation, i.e., to process data as they become available. To achieve this, the phase correction algorithm is split into multiple simultaneous tasks (Fig. 2): an edge trigger detects interferograms in the incoming data stream and their approximate occurrence time. It then forwards the 4 μs before (maintained using a buffer) and after the trigger to the interferogram analysis step. It digitally measures each incoming interferogram's phase, frequency (via a fast Fourier transform of the interferogram), and exact arrival time (via the first moment of the time-domain interferogram magnitude) and calculates fluctuations of the combs' carrier-envelope offset frequencies and repetition rates in less than 11 μs. This enables single-spectrum acquisition at up to 90 kHz rate.

Because phase correcting the current interferogram requires information about the preceding and the following interferograms, we delay processing data around the current interferogram until the next interferogram has arrived and been analyzed using a second buffer (located in external random access memory to support detunings as low as 2 Hz). Then, we correct each sample by a phase obtained via a linear interpolation between the phases of adjacent interferograms.

Lastly, offset frequency changes are corrected by resampling the final data stream such that interferograms occur at a constant rate. We calculate a resampling time grid from the measured interferogram arrival times of the last, current, and following interferogram and linearly interpolate the measured interferogram onto the resampling time grid. This step also removes possible sub-sampling-rate fluctuations of the interferogram arrival time. Due to the algorithm's streaming operation, we cannot send samples at a higher rate than the fabric clock (which occurs if the detuning increases). Therefore, we resample the data rate to 95% of the fabric clock, which allows us to correct for positive and negative detuning fluctuations of up to 5% at the cost of a marginal bandwidth reduction.

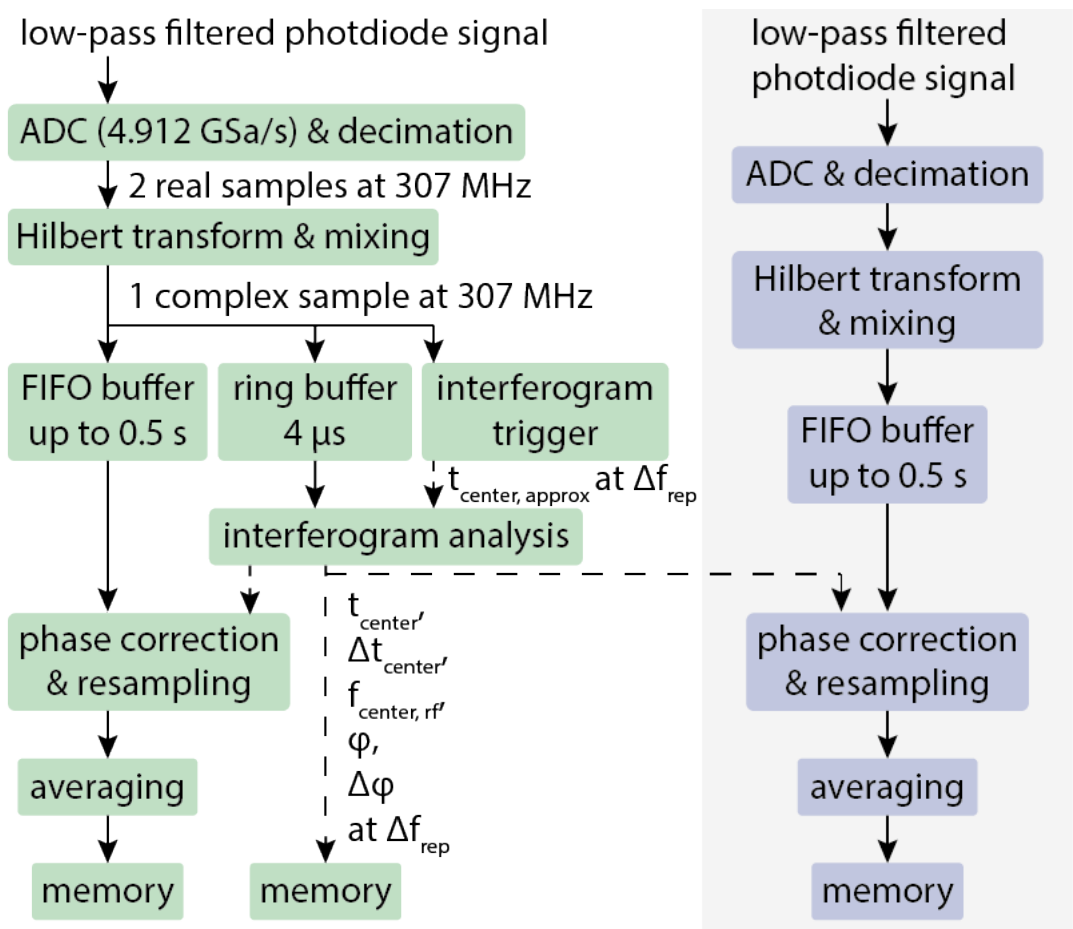

Fig. 2. **Schematic of the radio frequency system-on-chip-based self-correction algorithm.** Data moves through the field-programmable gate array along the solid arrows. Measured quantities move along the dashed arrows. All steps are executed concurrently until the data is handed off to the processing unit via direct memory access. The main (reference) channel (green) can be used independently or exploited to correct a second (signal) channel (blue).

Phase-corrected data is available after a latency of one offset period (e.g., 50 μs at 20 kHz offset frequency) + 11 μs and can be analyzed directly or coherently averaged. All steps, i.e., data acquisition, processing, recording, and display are implemented on a single radio frequency system-on-chip, reducing additional electronics to a photodiode and a low-pass filter. Correction data calculated from the main channel can be applied to a second channel with a software controllable delay and phase multiplier to compare a reference and signal channel, or to allow correcting weak or nonlinearly upconverted signals (see below).

## Comb Resolved Spectroscopy

To examine the viability of our real-time phase correction method, we first investigate the comb structure of our lasers at 525 nm wavelength (where we expect phase fluctuations to have a larger influence than at the fundamental wavelength): we evaluate unaveraged, continuous, 50-ms long photodiode signals containing 1000 interferograms. For visibility, we remove the radio frequency carrier of the interferograms.

Fig. 3a shows the signal before self-correction. Although the single-cavity laser only exhibits small phase drifts (i.e., changes from a real-dominated to an imaginary-dominated analytical signal), they considerably decrease the intensity of the recorded comb-resolved spectrum (see Fig. 3c) and smear out the lasers' comb structure (Fig. 3d and e). In comparison, Fig. 3b shows a corrected interferogram train recorded in close succession: after the algorithm compensates $\Delta f_{ceo}$ and $\Delta f_{rep}$ variations, the signal is always dominated by its real part.

Fig. 3c compares the uncorrected (red) and corrected (blue) comb-resolved spectrum, i.e., the Fourier transform of the data in Fig. 3a, b with the frequency axis adjusted for the down-conversion factor. Close-up views of the spectrum center (Fig. 3d) and the tail (Fig. 3e) after the phase correction reveal the individual comb lines with large amplitude spaced by $f_{rep}$ = 1 GHz. Looking at a single comb mode in the spectrum tail (Fig. 3f) highlights the quality of the correction: the comb mode exhibits a radio frequency linewidth of only 23 Hz, barely exceeding the Fourier limit (20 Hz) of a 50 ms-long trace (optical linewidth: 1.2 MHz).

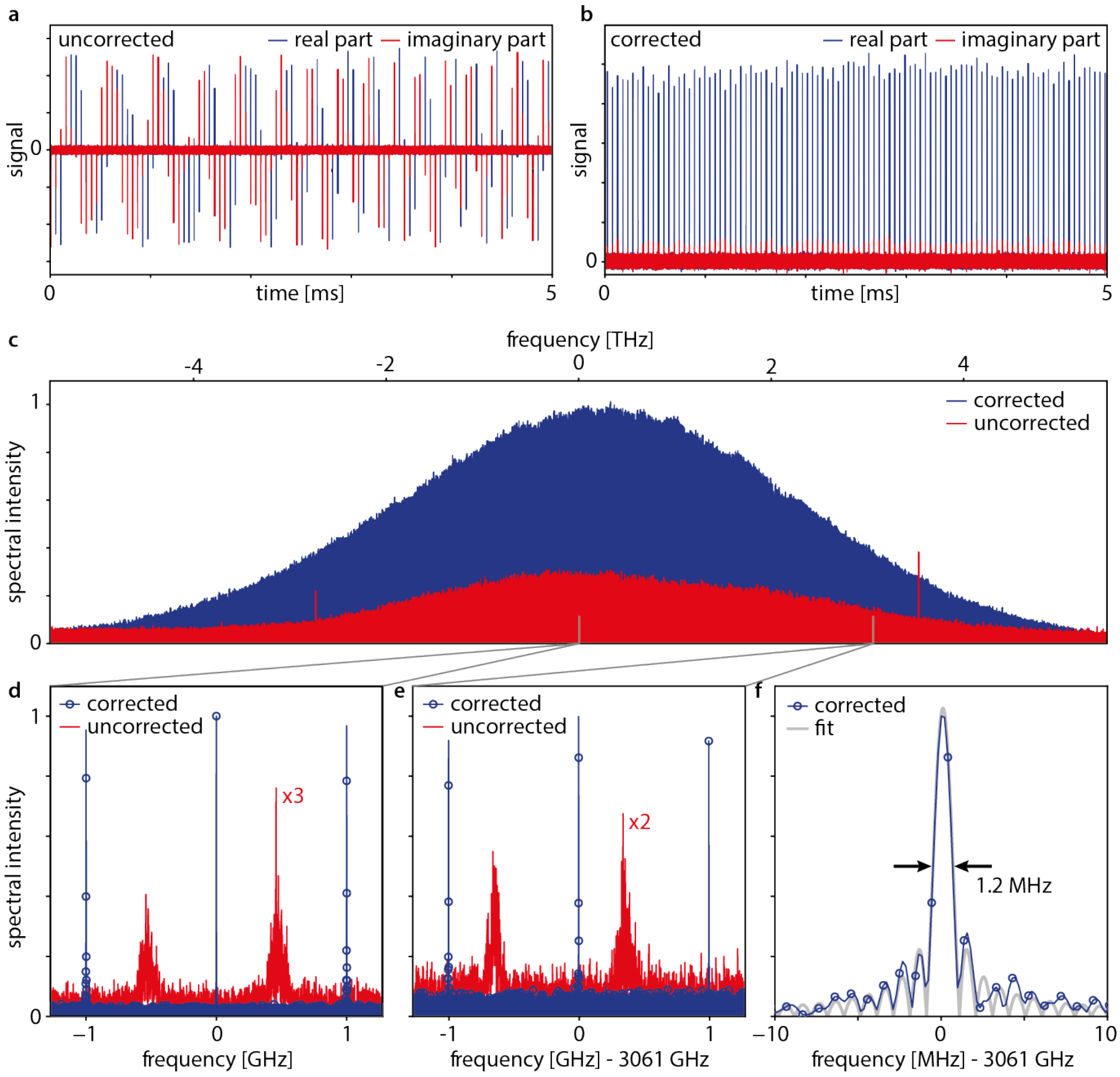

Fig. 3. **Corrected and uncorrected interferogram trains and comb resolved spectroscopy.** Comparison of the analytical signals calculated from the initial and corrected time-domain interferogram train with the radio frequency carrier frequency shifted to zero. a) The uncorrected signal slowly changes from mostly real-valued (blue) to mostly complex-valued (red), indicating that the phase of consecutive interferograms slips. b) The corrected signal is always dominated by its real part (blue), i.e., its phase is stable. c) The entire spectrum calculated from a 50-ms long measurement shows comb modes across more than 10 THz. The uncorrected spectrum (red) exhibits broad, low-intensity comb modes (the carrier frequency was set to zero). After correction, close-ups of d) the center and e) the tail of the spectrum demonstrate narrow comb modes with high amplitude across the laser bandwidth (the blue circles show data at the original resolution; the blue lines are calculated after zero-padding to mitigate aliasing). f) The corrected comb mode linewidth remains close to Fourier-limited even at the edge of the spectrum and matches the expected spectral shape (grey line: sinc function fit).

## Coherent Averaging and Spectroscopy

With the effectiveness of our spectrometer and phase correction algorithm proven, we can now coherently average interferograms in the time domain, allowing extended measurement times and thus high signal-to-noise ratios with minimal memory requirements. Fig. 4a shows two coherent averages of 20,000 interferograms: a reference without a sample (red) and a measurement after an iodine absorption cell (blue). By averaging over the individual interferograms, the free induction decay of iodine becomes visible with high fidelity even though it accounts for only 1 % of the overall signal strength. The free induction decay diminishes slowly and is present for the entire duration of an interferogram (50 μs for 20 kHz detuning). To examine the performance for longer averaging times, we determine the dependence of the signal-to-noise ratio of the reference on the number of coherently averaged interferograms, see Fig. 4b. The signal-to-noise ratio increases following a square root function and reaches ~2900 when averaging up to 3 million interferograms (150 s averaging time).

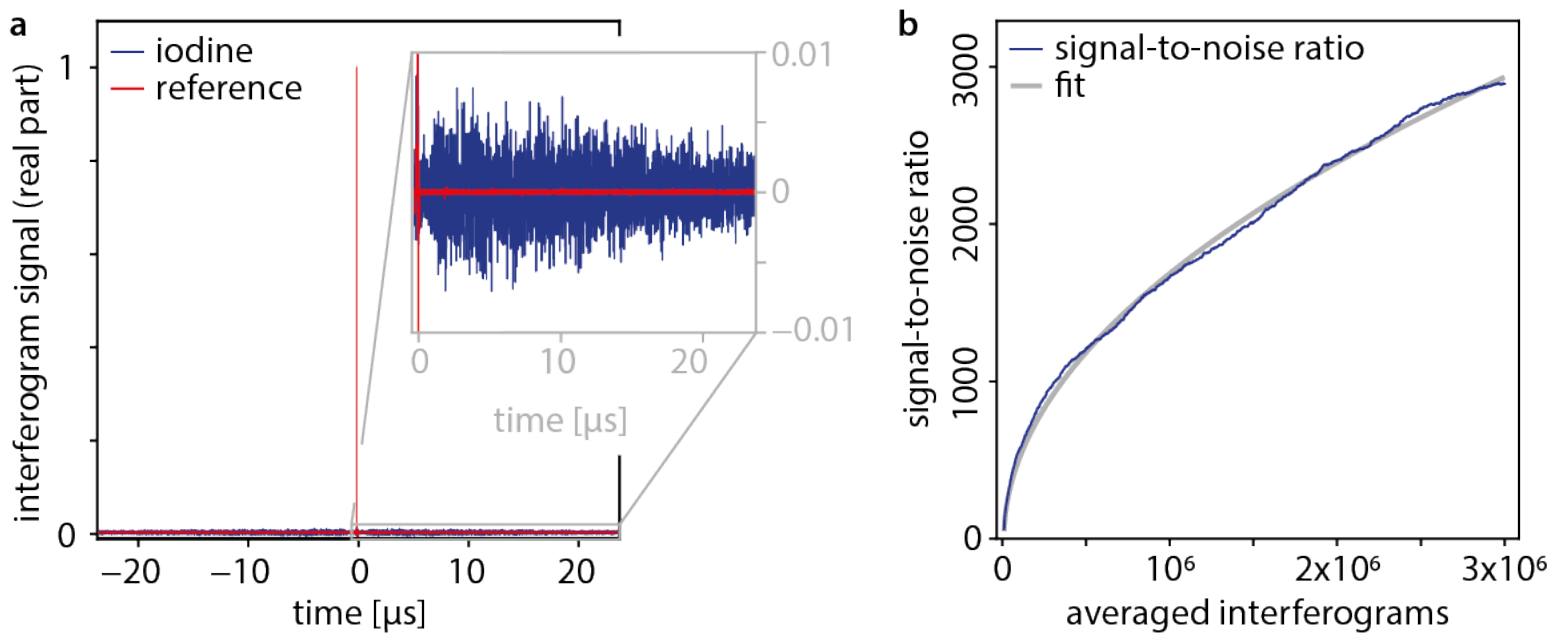

Fig. 4. **Time-domain coherent averaging.** a) The real part of 20,000 coherently averaged interferograms acquired without a sample (reference, red) and after an iodine cell (blue). Whereas the reference is dominated by a peak at the pulse overlap, zooming into the signal axis after the iodine cell reveals its free induction decay. b) Increasing the number of averaged interferograms increases the signal-to-noise ratio (blue) according to a square root function (grey).

Gaseous iodine is commonly used as a spectroscopic reference due to its narrow absorption features in the visible spectrum, originating from strong B-X rovibronic transitions[34]. To show that the phase correction algorithm works across the full bandwidth of our laser, Fig. 5 displays the spectrally resolved transmission of the iodine cell and close-ups of the low and high-frequency parts of the spectrum. We find excellent agreement between the measured iodine transmission and the literature[35] throughout the laser spectrum (12 THz). At room temperature, the Doppler width of iodine in this spectral region is ~100 MHz. As shown by the comb line width in Fig. 3f and the data in Fig. 5, the measured iodine spectra are only limited by the system's repetition rate and not by the real-time phase correction.

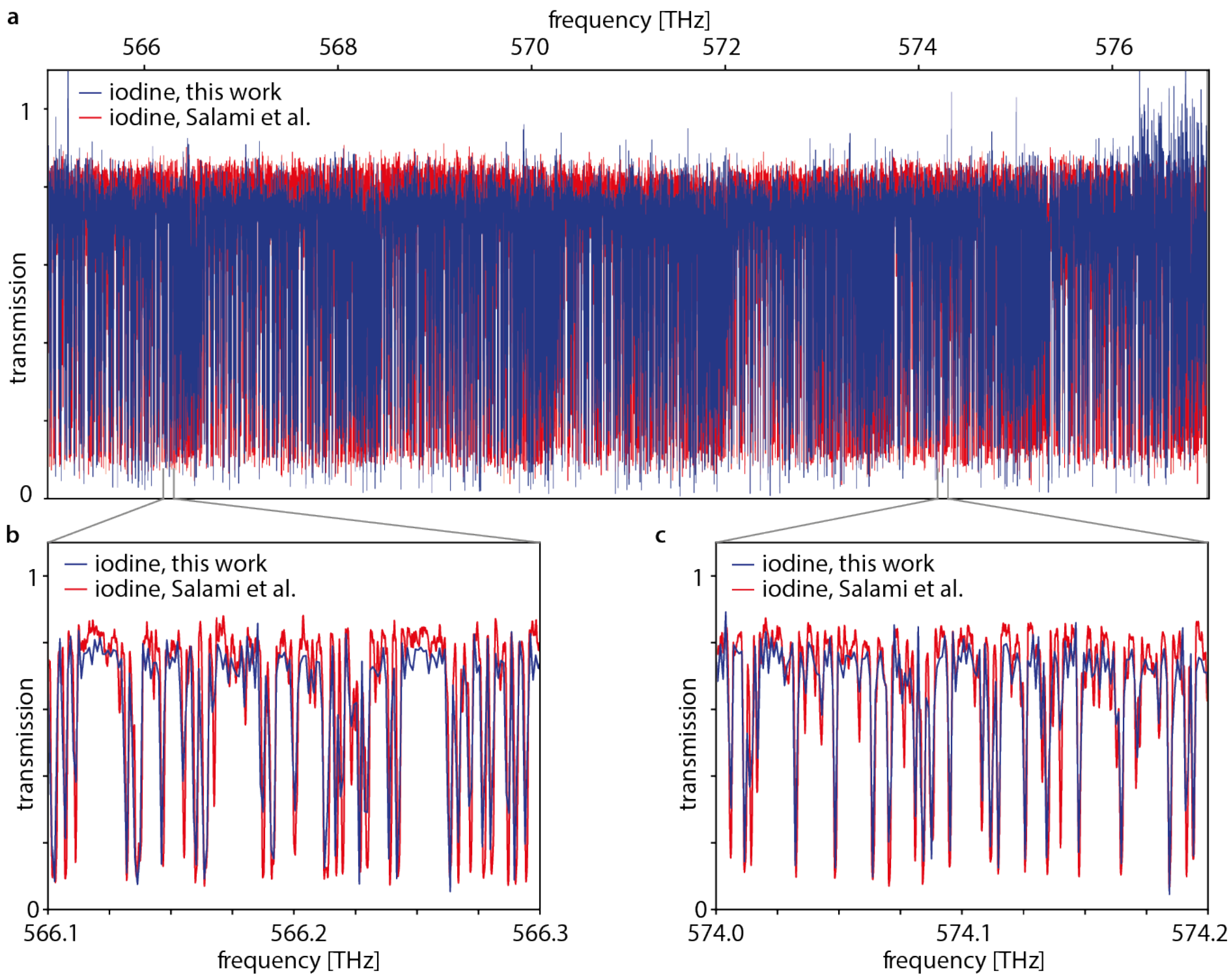

Fig. 5. **Measured iodine transmission (blue) compared with**[35] (red). a) Transmission across 12 THz bandwidth, calculated from the coherent average of 20,000 interferograms (measurement time: 1 s). Close-ups of the b) low- and c) high-frequency parts of the spectrum demonstrate the precision of our spectrometer (iodine cell length: 20 cm, pressure: 0.2 mbar; the transmission should not be compared quantitatively as not all lines are saturated by our cell).

Another common spectroscopic reference is the $3\nu_3$ vibrational band[36] of acetylene. The band contains fewer spectral lines than iodine, thus allowing a more straightforward comparison with a reference. To observe it, and show the adaptability of the spectrometer, we superimpose the near-infrared outputs of our combs in an acetylene-filled sample cell. Without changes, the phase correction algorithm coherently averages the interferograms generated by the dual-comb spectrometer and unveils the acetylene absorption in the transmitted spectrum (Fig. 6c). Because all steps - including averaging - are executed in a streaming fashion, we can implement long integration times and circumvent the memory constraints experienced by typical post-correction methods; Fig. 6 uses 50 s averaging time. After background subtraction, we find good agreement between the measured acetylene absorbance and its simulation (see Fig. 6a, b, experimental and simulation parameters: temperature: 300 K, acetylene/air ratio: 20 %, pressure: 220 mbar, interaction length: 105 cm) [37,38]. The small residuals over the 5 THz-broad vibration band highlight the resolving power of the free-running spectrometer (Fig. 6a top).

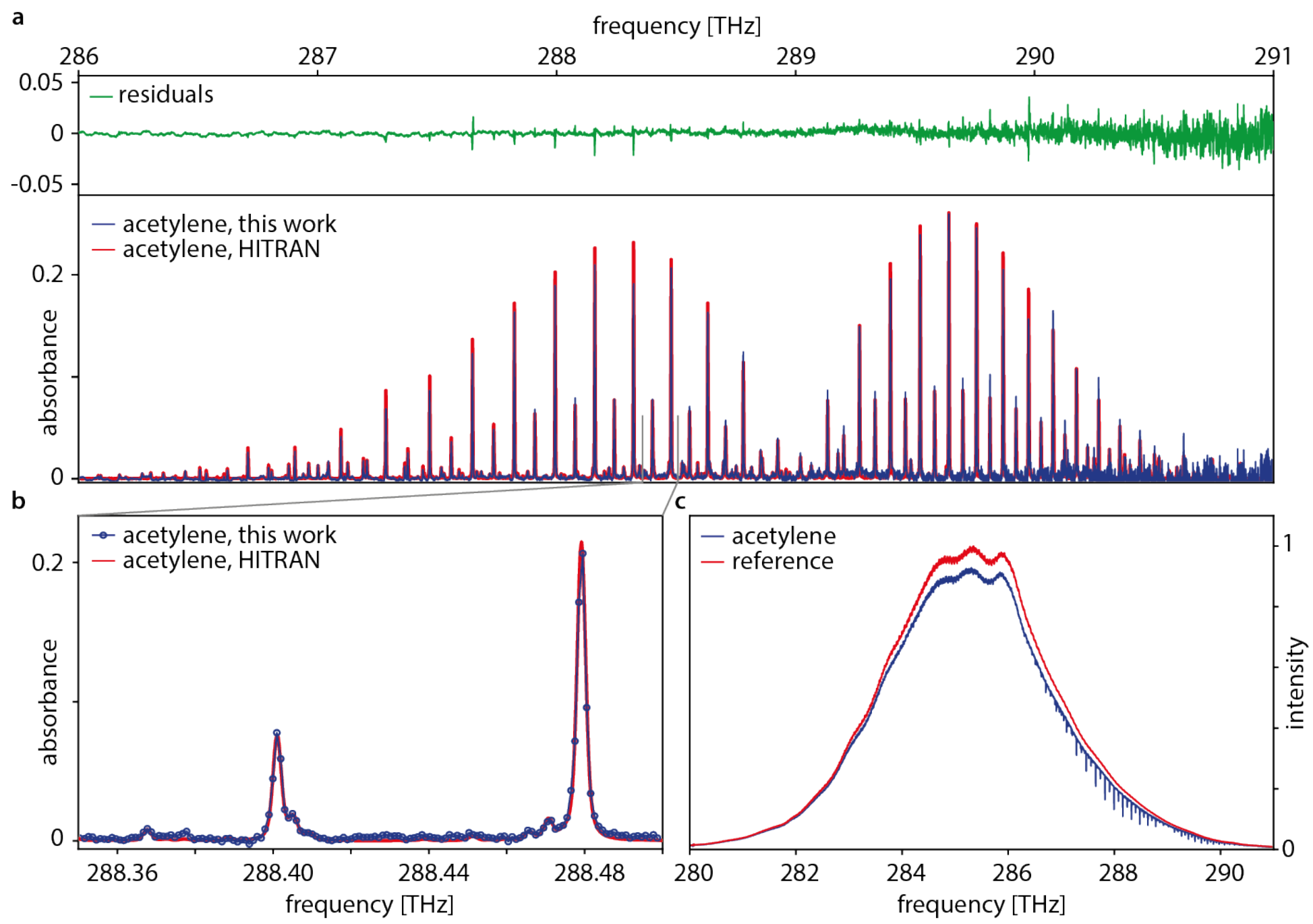

Fig. 6. **Acetylene absorbance measurements at 1050 nm wavelength.** a) Measured (blue) acetylene absorbance compared to a simulation (red)[37,38] and their difference (green). The absorbance across 5 THz bandwidth around the $3\nu_3$ vibrational band of acetylene centered at 289 THz [36], calculated from the coherent average of 1,000,000 interferograms (measurement time: 50 s, we subtract a linear baseline). b) A close-up of two double-peak structures. c) The measured laser intensity without (red) and with (blue) acetylene in the sample cell.

## Cross-Channel Correction

Dual-comb spectroscopy's high signal-to-noise ratio and short acquisition time make it attractive for field measurements. Nonlinear upconversion or spectral broadening of the combs' enables spectroscopy in new spectral regimes. However, these applications often hinder self-correction, as air turbulence or low signal levels may impede reliable detection of each interferogram and its phase. To still enable such measurements, our code supports a two-channel configuration: the combs' fluctuations are deduced from a reference channel and applied to correct a signal channel (see Fig. 2).

Cross-channel correction has – to our knowledge – been demonstrated using effortfully created reference signals from an f-to-2f interferometer and beating with a stable continuous-wave laser[39] or to the same interferogram train at different power to enhance the dynamic range[40]. Here, we show that one interferogram train's self-correction is sufficient to correct even a frequency-upconverted interferogram train with enhanced phase fluctuations.

To demonstrate the viability of our scheme, we use the setup depicted in Fig. 7a: two β-barium borate crystals generate weak second harmonic beams (~0.2 mW) from the fundamental laser outputs (1050 nm). Two photodiodes detect the fundamental (reference) and second harmonic (signal) interferogram trains. Our algorithm then calculates and corrects reference phase fluctuations, as detailed in Fig. 7b.

At the chosen laser settings, the phase of consecutive second harmonic (signal) interferograms changes twice as fast, complicating direct phase correction (Fig. 7c). However, a fixed relationship between the second harmonic and fundamental comb tooth frequencies ties their beat frequencies, $f_{rf,\ SHG} = 2\ (f_{optical}\ /\ m + \Delta f_{ceo})$ and their fluctuations to the reference interferogram train (here, $f_{optical}$ and $\Delta f_{ceo}$ are properties of the fundamental combs (see above)). Therefore, after applying a Hilbert transform, we digitally mix the signal arm with 307 MHz (twice the frequency used for the fundamental interferogram train), double the phase correction applied to the reference train, and apply it to the signal interferogram train. The corrected signal interferogram train exhibits a stable phase and can be coherently averaged (Fig. 7c).

We demonstrate this approach by revisiting iodine spectroscopy: we retain the cross-channel correction setup and introduce an iodine cell into the signal beam path. Fig. 7d shows the fundamental, second harmonic, and iodine-modified second harmonic spectra. Fig. 7e displays the iodine cell transmission in the same spectral range as shown in Fig. 5b. Although we do not directly measure the signal interferogram fluctuations, the transmission spectrum derived from the coherently averaged, cross-corrected interferograms closely matches both literature values [35] and our directly corrected measurement.

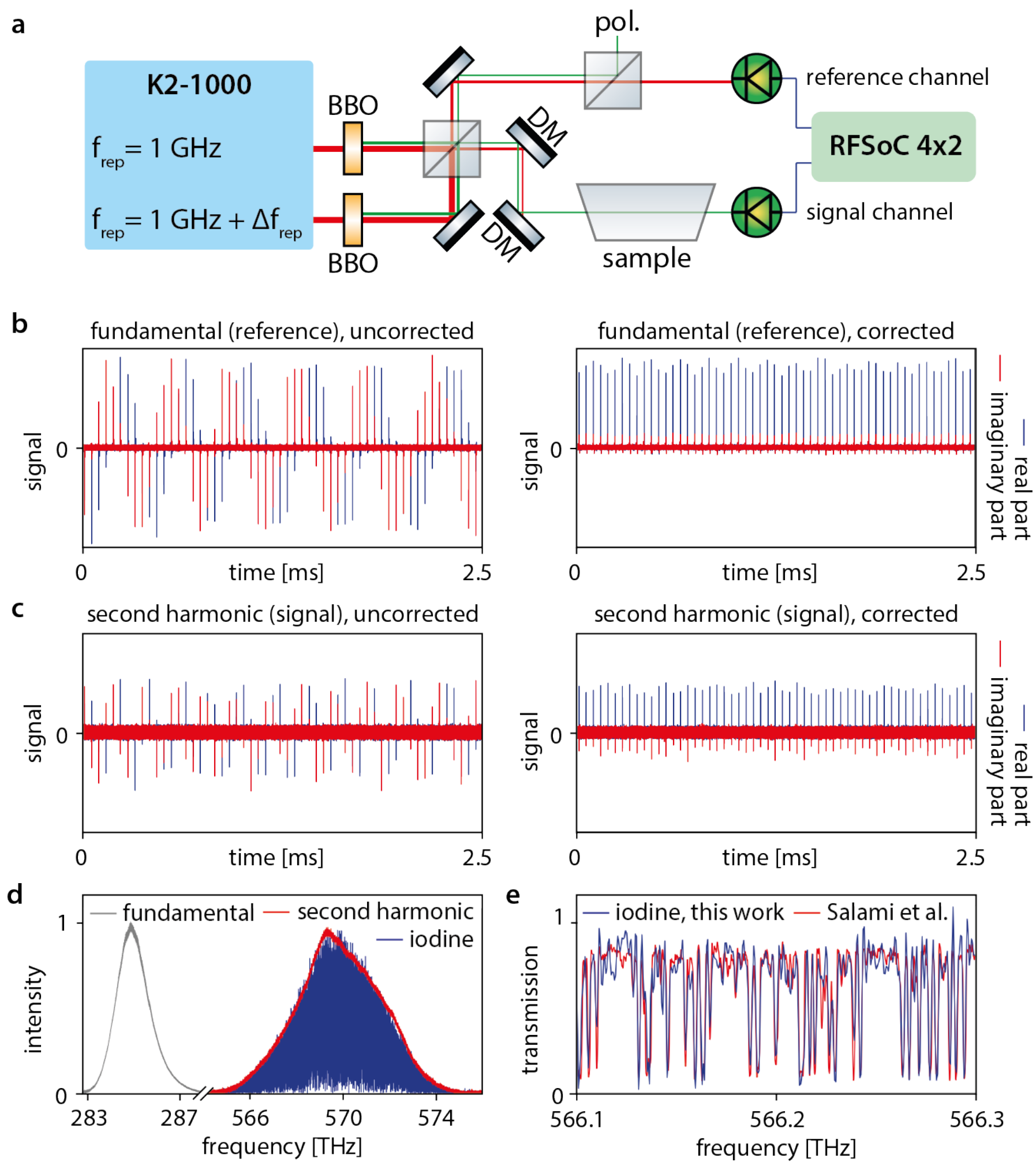

Fig. 7. **Cross-Channel Correction.** a) Setup: we collinearly create two weak second harmonic beams (green) using β-barium borate crystals (BBO) from two frequency combs (red) generated by a single-cavity dual-comb laser. We superimpose all beams using a non-polarizing beamsplitter. In the reference channel, a polarizing beamsplitter (pol.) isolates the fundamental beams whose beating is detected by a photodiode. In the signal channel, dichroic mirrors (DM) isolate the second harmonic beams whose beating is detected after a sample cell. b) Comparison of an initial (left) and a corrected (right) reference interferogram train. c) Comparison of an initial (left) and a corrected (right) signal interferogram train (without sample cell). All panels in b and c use the same ordinate scale. An inferior photodiode increases the noise floor and reduced second harmonic power decreases the signal level of c compared to b. As c is cross-corrected, neither influences the phase correction. d) The spectra of the reference (fundamental, grey), signal (second harmonic, red), and signal after the iodine-filled sample cell (iodine, blue), calculated from the coherent average of 20,000 interferograms (measurement time: 1 s). e) Measured iodine transmission (blue) compared with[35] (red) calculated from the spectra plotted in panel d (compare also with Fig. 5b and see Fig. 5 caption for iodine cell details).

## Conclusion

In conclusion, this work reports on a real-time self-correction algorithm for free-running dual-comb systems and its implementation to realize dual-comb spectroscopic streaming for everyday use. The field-programmable gate array-based code measures interferogram phase and repetition rate detuning deviations without external information from the interferogram train within 11 μs and corrects their effect in real-time.

Due to a large sampling bandwidth and fast processing, we support high repetition rates and detuning frequencies. This enables two interesting use cases: first, we can combine broadband sampling with fast acquisition times. Second, we can cross-correct low-quality interferograms even after nonlinear upconversion changes their optical and radio-frequency carrier.

We show that modern single-cavity dual-comb lasers fulfill the only prerequisite - a carrier-envelope offset frequency that drifts slowly compared to the detuning - making high-resolution, high signal-to-noise spectroscopy available to non-specialized laboratories and field applications. The presented system can independently acquire and coherently average broadband spectra to track fast chemical reactions or trace gas concentration changes with a kilohertz rate, reconstruct the comb-resolved spectrum of free-running frequency combs throughout their entire spectrum, or investigate weak absorption features with high fidelity.

## Code availability

The self-correction code is available at *https://github.com/marcus-o/real_time_rfsoc_phase_correction* .

## Data availability

The data presented in this paper is available at *https://doi.org/10.5281/zenodo.14899019* .

## Acknowledgments

We acknowledge the AMD University Program for software and hardware support. The authors gratefully acknowledge support from NAWI Graz. B.B. acknowledges funding from the European Union (ERC HORIZON EUROPE 947288 ELFIS). M.O. acknowledges funding from the European Union (grant agreement 101076933 EUVORAM). The views and opinions expressed are, however, those of the author(s) only and do not necessarily reflect those of the European Union or the European Research Council Executive Agency. Neither the European Union nor the granting authority can be held responsible for them.

## Author Contributions

Code: M. O.; Experimentation, Validation: A. E., C.G., M. S., B. B., M. O.; Funding: M.S., B.B., M.O.; Initial Draft: A. E., M. S., B. B., M. O.; All authors discussed the final manuscript.